# Small animal whole-body imaging with metamaterial-inspired RF coil

Mikhail Zubkov[1], Anna A. Hurshkainen[1], Ekaterina A. Brui[1], Stanislav B. Glybovski[1], Mikhail V. Gulyaev[2], Nikolai V. Anisimov[2], Dmitry V. Volkov[3], Yury A. Pirogov[4], Irina V. Melchakova[1]

[1]Department of Nanophotonics and Metamaterials, ITMO University, Saint-Petersburg, Russia

[2] Laboratory of Magnetic Resonance and Spectroscopy, Faculty of Fundamental Medicine, Lomonosov Moscow State University, Moscow, Russia

[3]Department of Physics of Accelerators and Radiation Medicine, Faculty of Physics, Lomonosov Moscow State University, Moscow, Russia

[4]Department of Photonics and Microwave Physics, Faculty of Physics, Lomonosov Moscow State University, Moscow, Russia

**List of abbreviations:**

RF – radiofrequency

FoV – field of view

SNR – signal to noise ratio

CNR – contrast to noise ratio

**Abstract**

Preclinical magnetic resonance imaging often requires the entire body of an animal to be imaged with sufficient quality. This is usually performed by combining regions scanned with small coils with high sensitivity or long scans using large coils with low sensitivity. Here, a metamaterial-inspired design employing a parallel array of wires operating on the principle of eigenmode hybridization is used to produce a small animal whole-body imaging coil. The coil field distribution responsible for the coil field of view and sensitivity is simulated in an electromagnetic simulation package and the coil geometrical parameters are optimized for the chosen application. A prototype coil is then manufactured and assembled using brass telescopic tubes and copper plates as distributed capacitance, its field distribution is measured experimentally using $B_1^+$ mapping technique and found to be in close correspondence with simulated results. The coil field distribution is found to be suitable for whole-body small animal imaging and coil image quality is compared with a number of commercially available coils by whole-body living mice scanning. Signal to noise measurements in living mice show outstanding coil performance compared to commercially available coils with large receptive fields, and rivaling performance compared to small receptive field and high-sensitivity coils. The coil is deemed suitable for whole-body small animal preclinical applications.

**Introduction**

Small animal imaging is crucial to a majority of preclinical research. A number of applications in preclinical magnetic resonance imaging (MRI) requires full body images of small animals (e.g., mice) to be acquired, for example angiography[1], fat quantification[2], contrast agent or drug delivery[3–5] and more[6]. Conventionally, such images are obtained either through stitching of multiple fields of view (FoV) sequentially acquired with small surface coil or with a single volume coil (usually, a birdcage coil). Both approaches have their own benefits and drawbacks.

Birdcage coils provide uniform excitation ($B_1^+$) and reception ($B_1^-$) within their internal volume[7]. The areas of field homogeneity provided by such coils are enough for small animal whole-body imaging even at high frequencies where the wavelength of the used radiofrequency (RF) field shortens leading to uniformity issues in human MRI[8]. The downside of using volume coils is in the low sensitivity they provide and the large area they collect the noise from. These two combined effects lead to generally low signal-to-noise ratio (SNR) in images obtained with birdcage coils. Another drawback of a birdcage is its closed geometry completely surrounding the studied subject except for the front and the back coil ends. In applications where an external excitation or monitoring is required (e.g. acoustic excitation[9]), this coil design provides limited access and is, therefore, inconvenient.

The other option for full-body imaging, i.e. using small surface coils and combining the results during post-processing, provides high SNR due to better sensitivity of small surface coils at depths comparable to coil size. The small volume coverage of surface coil is corrected for during post-processing by stitching a number of fields of view together. As often image stitching is performed manually or semi-automatically[10,11] imaging *in vivo* introduces a risk inconsistencies between the registered volumes due to physiological movement. Another aspect of small surface coils that is usually not corrected during post-processing is the non-uniformity of $B_1^+$ and $B_1^-$ leading to uneven signal (and consequently SNR and CNR) distribution in the acquired images, although robust algorithms for image fusion might provide seamless results even when the original images have non-uniform sensitivity[12].

There is a gap in technological solutions for this sensitivity versus coverage tradeoff, which this work is aimed to close. An intermediately sized coil with $B_1^+$ and $B_1^-$ uniformity better than one of conventional surface coils and rivaling sensitivity should provide whole-body small animal imaging without the need of post-acquisition fusion as well as SNR high enough to allow preclinical MRI experiments to be carried out.

## Experimental

### Coil design

In order to perform full-body imaging with high SNR and resolution a metamaterial-inspired RF coil (Figure 1) with a parallel wire array geometry was assembled for a 300 MHz MR-scanner. The coil design was inspired by so-called 'mushroom' metamaterial structures, which are subwavelength periodic arrays of wires loaded by square capacitive metal patches forming an artificial magnetic conductor[13]. The operational principle of the coil is based on a hybridization of eigenmodes in an array of parallel non-magnetic wires[14]. The coil comprises wire resonator inductively coupled with small non-resonant magnetic loop.

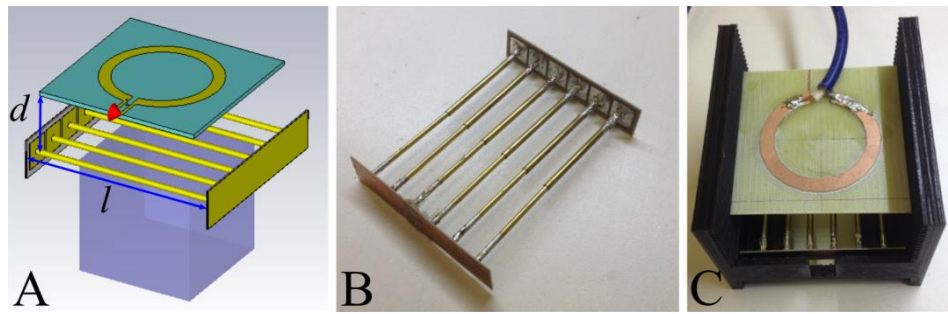

**Figure 1 Radiofrequency coil for $^1$H 7 T small-animal imaging: (A) simulation model with optimized dimensions *d* and *l*; (B) manufactured wire resonator of the coil; (C) assembled loop coil, resonator and holder.**

Resonator has subwavelength dimensions in all directions due to the attached distributed capacitance[15]: six brass tubes are connected at both ends to rectangular copper patches deposited on a high-quality low-loss dielectric 0.508 mm thick substrate Arlon 25N (with $\varepsilon = 3.38$ and $\tan \delta = 0.0025$ at 10 GHz). The common ground plane at the opposite side of the substrate provides a capacitive interconnection of all tubes. It has been shown that in such resonator type multiple surface eigenmodes can be excited all having different $B_1^+$ patterns. The fundamental eigenmode has the most homogeneous $B_1^+$ distribution with the highest penetration depth into the subject[14]. Inductive coupling of the resonator with the feeding loop provides excitation of the first eigenmode of the resonator both in transmission and reception regimes. Tuning the first eigenmode of the resonator to the Larmor frequency of $^1$H at 7 T (300.8 MHz) is carried out by changing the length of the tubes, in a range from 57 mm to 80 mm, while the value of capacitance is kept constant due to the fixed size of the capacitive patches (9×9.5 mm$^2$). Matching of the coil is performed through selection of optimal coupling between the resonator and the feeding loop, which can be varied by modifying the distance between the loop and the resonator.

The radiofrequency coil design was simulated in the commercial software CST Microwave Studio 2016 (Computer Simulation Technology GmbH, Germany) using the Frequency Domain solver in the presence of the RF shield model (i.e., a perfect conductive tube with the inner diameter of 200 mm) and the homogeneous cubic phantom (40×40×40 mm$^3$) with a dielectric permittivity of 78.4 (Figure 1, A). Length of the tubes (*l*) as well as the distance between the resonator and a feeding loop (*d*) was optimized in simulations aiming to perfectly tune and match the coil at the resonant frequency of 300.8 MHz.

The metamaterial-inspired radiofrequency coil was then manufactured using the optimized parameters of the numerical model. Six telescopic brass tubes were soldered at both ends to the patches of two printed circuit boards representing the constructive distributed capacity (Figure 1, B). 40-mm diameter feeding loop was implemented on a 1.5 mm FR-4 dielectric substrate. Connection of the feeding loop with the transceiver of the MRI scanner was provided by the coaxial cable. The resonator with the feeding loop were attached to a 3D-printed dielectric holder (Figure 1, C). The manufactured coil tuning and matching range was tested outside of the scanner using PNA E8362C vector network analyzer (Agilent Technologies Inc, USA), afterwards the coil was subject to scanning tests and used in MRI experiments as discussed in the next sections.

**Field homogeneity measurements**

Verification procedures for coil specifications and field distribution were carried out on a 7 T Bruker BioSpec 70/30 USR (Bruker BioSpin GmbH, Germany) horizontal 200 mm bore scanner running the ParaVision 5.0 software. These experiments was carried out in the Centre for Collective Usage "Biospectrotomography" supported by the Faculty of Fundamental Medicine of Lomonosov Moscow State University, Moscow, Russia. The coil RF field distribution was measured by scanning a homogeneous rectangular phantom (40×40×40 mm$^3$) filled with distilled water with 0.1 ml Magnevist (with the solution measured spin-lattice relaxation time $T_1 = 40$ ms) encompassing a large portion of the coil desired field of view. The coil was positioned horizontally above the phantom with wires pointing along the main magnetic field $B_0$. A number of gradient echo images of the phantom were acquired while changing the RF power applied to the coil by varying the pulse nominal flip angle and keeping the pulse duration and reference attenuation constant. The $B_1^+$ field maps were acquired with the 2D sequences with isotropic 1×1×1 mm$^3$ resolution, 64×64×41 mm$^3$ field of view encompassing the whole scanned phantom, TE = 7 ms and TR = 1000 ms. The acquired data was fitted with the sine function with the resulting fitted amplitude providing the $B_1^-$ field distribution and the fitted frequency – the $B_1^+$ distribution[16].

*In vivo* **imaging**

Having measured and analyzed the field distribution of the new coil and having found it being suitable for performing whole-body imaging *in vivo* scans were performed on the aforementioned scanner. *In vivo* images of 2 outbred mice BALB/c of age 5-6 months and weight 35-40g were acquired to test the coil whole-body imaging performance. Experimental procedures were conducted in accordance with the European Community Council directives 2010/63/EU and were approved by the local institutional animal ethics committee. In order to reduce body motion artifacts the mice were provided with isoflurane via face mask. Images were obtained in consecutive experiments with two types of commercially available volume coils: RF

RES 300 1H/13C T10334 (inner diameter of 72 mm) and RF RES 1H T6594 (inner diameter of 115 mm). Additionally a small FoV image of mouse kidneys was acquired with a rat-brain receive-only surface coil (T11205 with length, width and height of 123 mm, 64 mm, 31 mm respectively with a resonator length of 17 mm) with excitation provided by the smaller birdcage coil (i.e., the cross-coil operating mode). In order to provide comparable imaging conditions care was taken to ensure proper tuning of the coils, automatic procedures for transmitter power calibration were checked to provide proper power attenuation values, receiver gain was kept constant and $B_0$ shimming was automatically optimized for each coil and mouse position.

Full body imaging was performed with a 2D SE pulse sequence using the following parameters: FoV = 40×100 mm², matrix = 200×200, slice thickness = 1.5 mm, flip angle= 90°, refocusing flip angle = 180°, TR = 800 ms and TE = 14 ms. Small FoV kidney imaging was performed with a similar pulse sequence with the parameters: FoV = 40×30 mm², matrix = 266×200, slice thickness = 1 mm, flip angle= 90°, refocusing flip angle = 180°, TR = 800 ms and TE = 13.64 ms with the exception of cross-coil imaging, where TE = 14.34 ms was used due to hardware constrains.

## Results

### Coil design and simulations

The coil length ($l$) and distance to the loop coil ($d$) was optimized to achieve the best tuning at the spectrometer frequency, resulting in the optimum parameters $l$ = 65 mm $d$ = 24 mm. The frequency dependence of the reflection coefficient of the simulated coil with optimized parameters and a similar measured dependence are depicted in Figure 2, showing proper tuning and matching of the coil with the reflection coefficient lower than -19 dB at the scanner operating frequency.

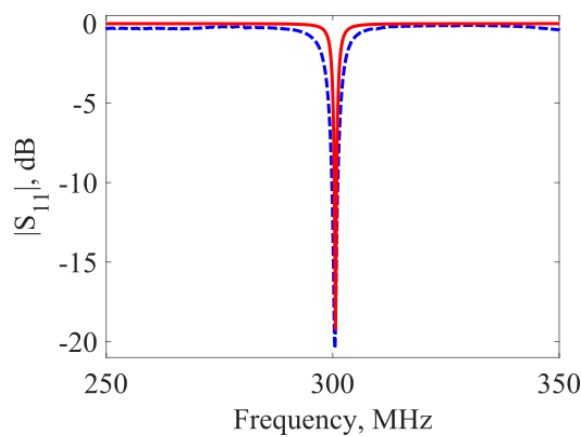

**Figure 2 Simulated (-) and measured (--) frequency dependencies of the reflection coefficient of the radiofrequency coil with optimal wire length *l* and wire separation *a*. The difference in the measured and simulated Q-factor (resulting in the change in the resonance line width) is due to the absence of the RF shield in on-bench measurements.**

The $B_1^+$ field distribution created by the coil was simulated on the optimized model to assess the field homogeneity, to find areas for possible small animal placement and to be later compared to the experimentally obtained field profiles. Calculated field distributions in the vicinity of the coil and inside the phantom volume at the top of the phantom (coronal plane) and in the central cross-section (sagittal and axial planes) were extracted from a larger simulated volume (Figure 3, D, E and F, for coronal, sagittal and axial planes correspondingly).

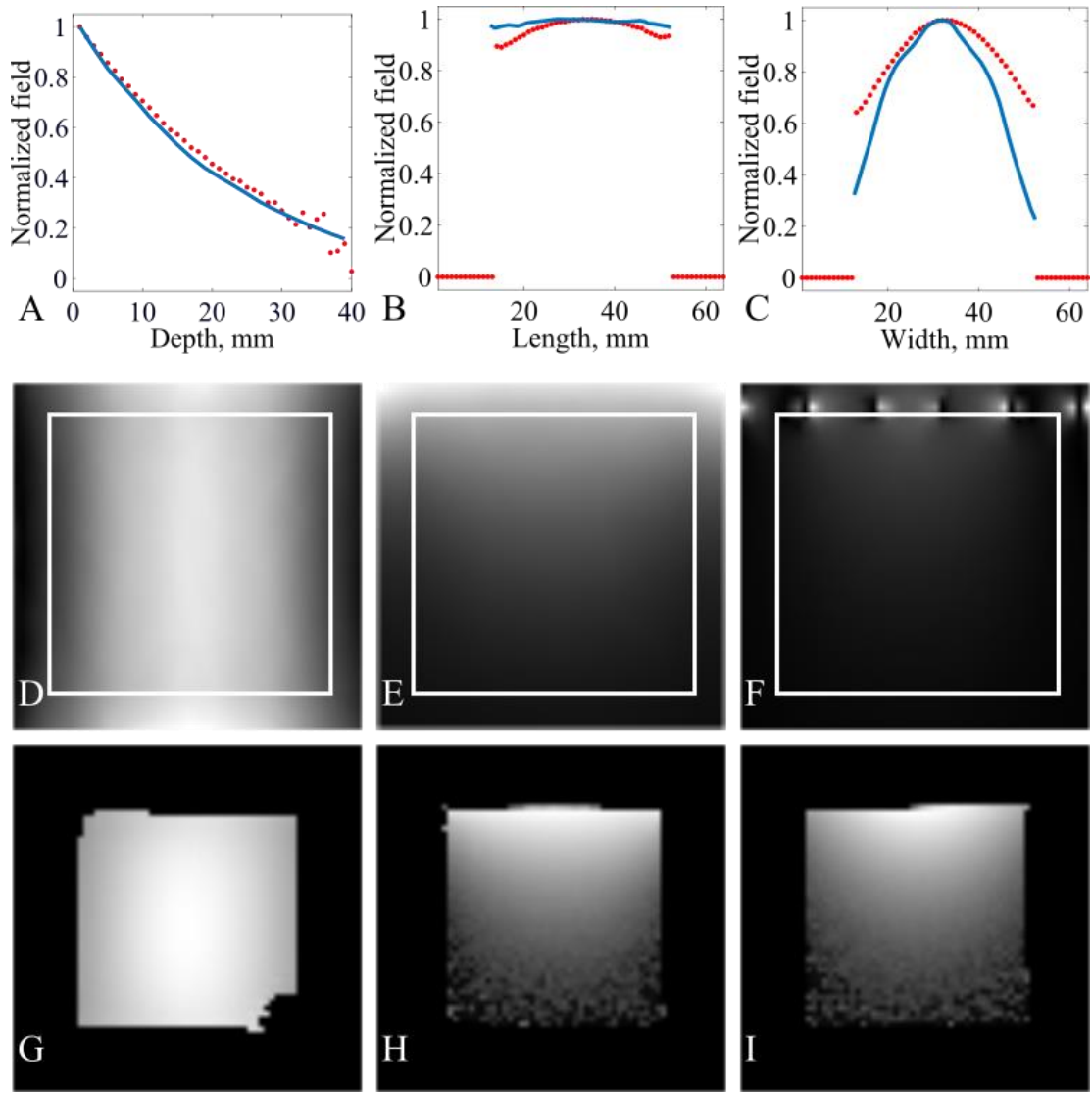

**Figure 3 $B_1^+$ metrics of the metamaterial-inspired coil. (A) – simulated (–) and experimental (·) field profile in the center of the 40×40×40 mm³ homogeneous phantom normally to the coil wire plane. (B) – field profiles in the direction of the wires 1 mm inside the phantom, (C) – field profiles in the direction across the wires. (D) – simulated map of the normalized RF field in the coronal plane 1 mm inside the phantom (the white box shows the phantom border), (E) – in the sagittal plane in the center of the phantom, (F) – in the axial plane. (G-I) – $B_1^+$ field distribution maps as measured on a rectangular 40×40×40 mm³ homogeneous phantom. (G) – coronal plane 1 mm inside the phantom, (H) and (I) – sagittal and axial planes in the in the center of the phantom.**

In order to facilitate the comparison between the experimentally measured field and simulation prediction field profiles inside the phantom, i.e. the dependencies of the $B_1^+$ on the depth, length and width of the phantom, were extracted from the simulation. The profiles correspond to the field distributions along the central axis of the phantom for depth and to the top surface of the phantom for length and width (Figure 3, A, B and C correspondingly). The field profile variation (standard deviation over mean) inside the phantom along the wires was found to be on the order of 1%, the variation across the wires was substantially higher, 33%.

## Field homogeneity measurements

In order to compare the simulation results to the experimentally obtained coil field distribution maps, the latter were normalized to the maximum field measured in the phantom volume (Figure 3). As expected, the coil was performing better closer to the coil surface with the transmission and reception efficiency diminishing with distance from the wire plane. The field drop off with distance, being the property limiting the use of surface coil for whole-body imaging was of particular interest. The comparison between the simulated field of the coil and the experimentally acquired $B_1^+$ distribution (Figure 3, A) shows fine correspondence in field drop off. A significant sensitivity was found to be retained at the planned mouse positioning distance (at least 40% of the sensitivity of the coil remained 20 mm inside the phantom in the planned ventral-dorsal direction).

Both simulations (Figure 3, D-F) and experimental maps (Figure 3, G-I) show the presence of a homogeneous field region suitable for positioning a small animal along the wire direction up to 20 mm away from the wire plane. Field variation profiles along selected directions were extracted from the field map and compared to similar simulation results. The experimental profiles show $B_1^+$ field distribution similar to one predicted in simulations (Figure 3, A-C). As predicted by simulations, field profile along the wires (Figure 3, B) shows better uniformity in the proposed cranial-caudal direction than in left to right direction (Figure 3, C): 3% variation across the 40 mm of the phantom in cranial-caudal direction versus 13% variation across the 40 mm of the phantom in left-right direction (across the wires).

## *In vivo* imaging

Field measurements and simulations have shown that whole-body mouse imaging with the proposed coil will be possible for proper mouse positioning. During the *in vivo* imaging experiments mouse torso was properly visualized in the chosen position. According to the simulations and field measurements the image quality was expected to diminish over the distance from the wire plane. The slice images on different distances from the coil (Figure 4) show sufficient image quality with the SNR of the images obtained with the resonator coil only slightly higher than one provided by the smaller birdcage coil on the slices closest (dorsal coronal slices) and furthest (ventral coronal slices) from the wire plane and surpassing it by around 300% in the area of optimal coil transmission and reception (Table 1).

Comparing small FoV images provided by birdcage coil (Figure 5, B and C), surface coil (Figure 6, B) and new resonator coil (Figure 5, A and Figure 6, A) shows renal pelvis and calyces as well as larger renal vessels being equally well visualized on both new coil and surface coil images, with the majority of the structures being obscured by noise in the both of the birdcage coil image series. Comparison of SNR in small FoV imaging (Table 2) shows an over 200% increase in SNR with the resonator coil compared to both birdcage coils and rivaling SNR with the surface coil (with the full-body resonator coil showing only 25% less SNR compared to the surface coil).

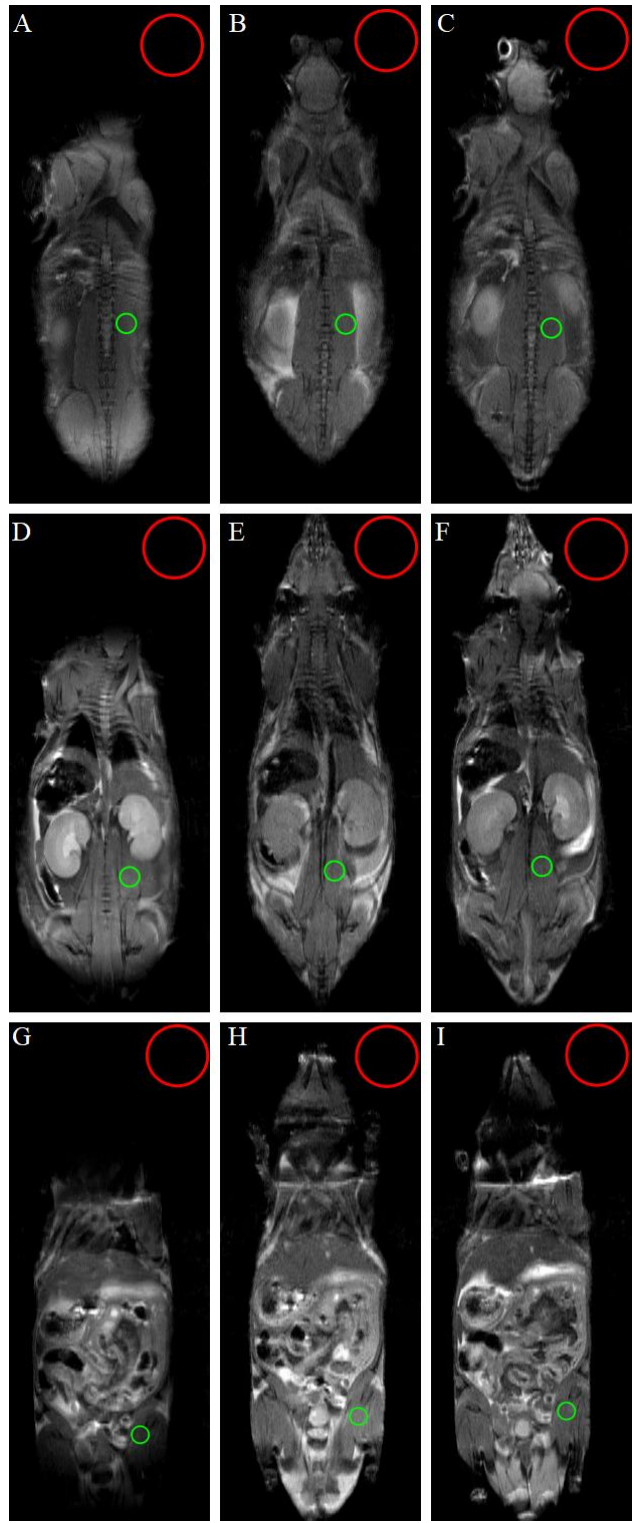

**Figure 4** Whole-body mouse images at (A-C) spine level (closest to the resonator coil), (D - F) kidney level (intermediate distance) and (G – I) liver level (furthest from the resonator coil). Images A, D, G were acquired with the proposed resonator coil, images B, E, H – with the medium 72 mm T10334 birdcage coil and images C, F, I- with the larger 115 mm T6594 birdcage coil. Images are presented at the same window/level settings. Regions of interest for SNR calculation are shown by green (muscle region for signal calculation) and red (signal-free region for noise calculation) circles. A summary of SNR values at selected RF field penetration depths is presented in Table 1.

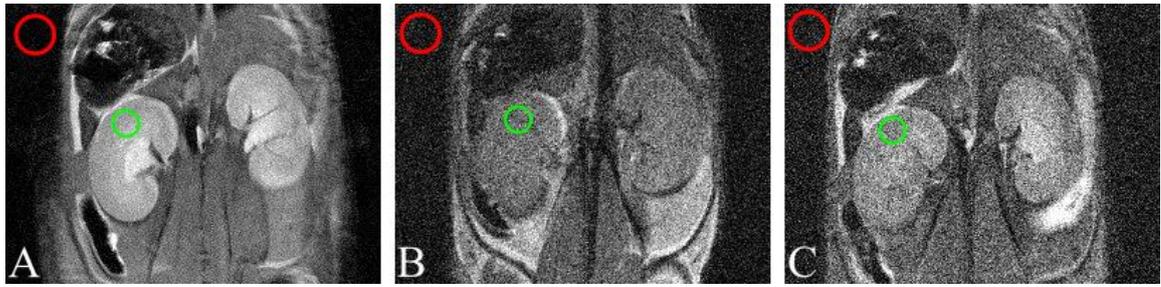

**Figure 5** Mouse kidneys imaged with A – new resonator coil, B - 72 mm T10334 birdcage coil and C – with the larger 115 mm T6594 birdcage coil. Images are presented at the same window/level settings. Regions of interest for SNR calculation are shown by green (kidney parenchyma region for signal calculation) and red (signal-free region for noise calculation) circles. A summary of SNR values is presented in Table 2.

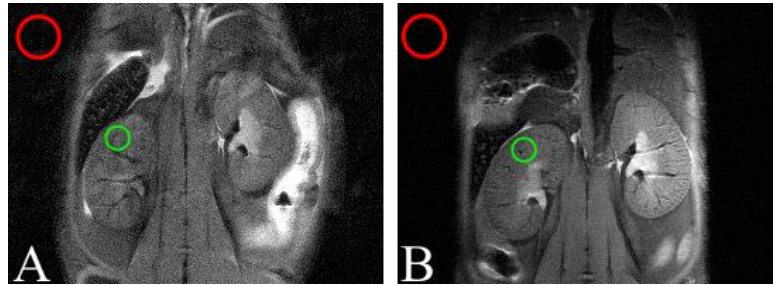

**Figure 6** Second mouse kidney images obtained in a different scanning session (compared to Figure 4 and Figure 5 images). Image A was obtained with the proposed resonator coil, image B – with a rat-brain receive-only surface coil T11205. Images are normalized to the signal value in muscle and presented at the same window/level settings. Regions of interest for SNR calculation are shown by green (kidney parenchyma region for signal calculation) and red (signal-free region for noise calculation) circles. A summary of SNR values is presented in Table 2.

**Table 1** SNR measurements summary for whole-body imaging at different distance from the resonator coil surface compared to SNR measurements in images from birdcage coils acquired at the same level. Areas for signal and noise measurements have been chosen as highlighted in Figure 4.

|  | Resonator coil | 72 mm birdcage | 115 mm birdcage |
| --- | --- | --- | --- |
| Dorsal | 52 | 25 | 23 |
| Median | 102 | 29 | 25 |
| Ventral | 67 | 37 | 28 |

**Table 2** SNR measurements summary for kidney imaging with the proposed metamaterial-inspired resonator coil compared to SNR measurements in images acquired with birdcage coils and surface coil. The positioning of regions of interest for signal and noise calculation is shown in Figure 5 and Figure 6.

|  | Mouse 1 | Mouse 2 |
| --- | --- | --- |
| Resonator coil | 17.4 | 12.8 |
| Birdcage coil, 72 mm | 5.8 | - |
| Birdcage coil, 115 mm | 6.3 | - |
| Surface coil | - | 17.0 |

## Discussion

The tested metamaterial inspired resonator coil design simulations have shown a field distribution with high potential for whole body mouse imaging with high SNR and large field of view. The measured RF field distribution of the real RF coil showed fine correspondence with the simulated results. The RF field distribution along the coil wires (corresponding to the proposed cranial-caudal direction of mouse positioning in the coil) showed high homogeneity in phantom imaging (3% variation of the RF field magnitude across the 40 mm of the phantom). The field distribution across the wires was less homogeneous (13% field variation across 40 mm of the phantom in the proposed left to right direction of mouse imaging). The field distribution in left to right direction on the other hand has a visible maximum in the center of the imaging field well suited for placing a small animal in the optimum reception and transmission field area. The coil field rapidly decreases normally to the wire plane, where the dimensions of the imaged animal should be the smallest (i.e., ventral-dorsal direction). On the 20 mm of common mouse thickness in ventral-dorsal direction the coil field drop-off was below 60% resulting in SNR still higher than the volume birdcage coils can provide. The inhomogeneous field distribution normally to the coil resulted in inhomogeneous flip angle distribution in the said direction, which in turn (when classical pulse calibration via finding an echo minimum in an axial slice[17] was used) resulted in optimum flip angle being only tuned in a limited spatial region. This can be seen in the *in vivo* SNR measurements (Table 1), where greater field in dorsal regions of the mouse (meaning greater sensitivity) provides less SNR due to flip angle being larger than optimal angle for the selected pulse sequence. Similar SNR variation was observed in small FoV imaging. Maximum SNR was nevertheless provided by the coil in the visceral region of the imaged animal, which is usually of greater interest than the peripheral regions of the animal. Moreover such field and sensitivity distribution allows obtaining maximum SNR at desired field penetration depth by tuning the RF power to the coil if different portions of the object present significant interest.

The SNR measurements *in vivo* show the metamaterial inspired coil sensitivity surpassing one of volume birdcage coils by at least 200% in the maximum sensitivity region and by at least 80% off the region of the highest coil sensitivity. The coil thus shows exceptional performance in its designated task (i.e., whole-body mouse imaging). Additionally comparing it to a number of coils in small FoV application has shown it to provide sensitivity high enough to visualize structures attainable only through the use of small receive-only surface coils. This allows the coil to be used in multiresolution applications, where both a large field of view and a detailed image of particular region are desirable (e.g., drug delivery monitoring). One of the downsides of the coil in current implementation is a limited field of view in cranial-caudal direction, resulting in the mouse head and the most caudal regions being impossible to image simultaneously. Nevertheless it was possible to image the head and partially the body of the mouse simultaneously by moving the coil cranially. In that case, the body was imaged approximately down to the mouse urinary bladder level with the more caudal structures being outside the coil FoV. In case full body coverage is required the coil design should be modified by reducing the structural capacity and extending the wire length, providing therefore a longer FoV at the same resonant frequency. Further improvements can include an active detuning circuit to operate the coil in receive-only mode with excitation by volume coil (if delivering sufficient pulse power to the coil is possible) thus providing uniform flip angle across the imaging volume and high receive sensitivity in whole-body applications.

The tested metamaterial-inspired coil has therefore been shown to be suitable for whole-body mouse imaging with high SNR, capable of providing large and small FoV images without mouse or coil repositioning and having potential for further improvements if a need for better transmit homogeneity or larger FoV arises. At the same time the tested coil geometry is open allowing easier access to the animal for suitable applications.

## Acknowledgements

This work was financially supported by the Ministry of Education and Science of the Russian Federation (Zadanie No. 3.2465.2017/4.6). This project has also received funding from the European Union's Horizon 2020 research and innovation programme under grant agreement No 736937. Additionally, the authors would like to thank Dr. Redha Abdeddaim for useful discussions.